\documentclass[conference]{IEEEtran}
\IEEEoverridecommandlockouts
\usepackage{cite}
\usepackage{amsmath,amssymb,amsfonts}
\usepackage{algorithmic}
\usepackage{graphicx}
\usepackage{textcomp}
\usepackage{xcolor}
\usepackage{amsthm}
\usepackage{algorithm}
\usepackage{stfloats}
\usepackage[colorlinks=true,
            linkcolor=blue,
            anchorcolor=blue,
            citecolor=blue,
            bookmarks=false]{hyperref}
\usepackage[caption=false,font=normalsize,labelfont=footnotesize,textfont=sf]{subfig}
\usepackage{epstopdf}
\def\BibTeX{{\rm B\kern-.05em{\sc i\kern-.025em b}\kern-.08em
    T\kern-.1667em\lower.7ex\hbox{E}\kern-.125emX}}
\begin{document}

\title{Joint Hybrid Beamforming and Trajectory Design for Multi-UAV-Enabled Cell-Free Multi-Static ISAC}

\author{
    \IEEEauthorblockN{
        Chen Chaoran\IEEEauthorrefmark{1}, 
        Zhang Yuhao\IEEEauthorrefmark{1}, 
        Pan Zhiwen\IEEEauthorrefmark{1}\IEEEauthorrefmark{2}, 
        and Liu Nan\IEEEauthorrefmark{1}
    }
    \IEEEauthorblockA{
        \IEEEauthorrefmark{1}National Mobile Communications Research Laboratory, Southeast University, Nanjing, China
    }
    \IEEEauthorblockA{
        \IEEEauthorrefmark{2}Purple Mountain Laboratories, Nanjing, China
    }
}

\maketitle

\begin{abstract}
This paper investigates a joint hybrid digital-analog beamforming and trajectory design for a cell-free multi-static integrated sensing and communication (ISAC) system supported by multiple unmanned aerial vehicles (UAVs).
Specifically, these UAVs cooperatively serve ground users and perform multi-static sensing to detect the target.
We formulate a weighted sum-rate (WSR) maximization problem by jointly optimizing the hybrid beamformers and the UAV trajectories.
This joint design explicitly accounts for practical constraints, including transmit power budgets, sensing signal-to-noise ratio (SNR) requirements, UAV kinematic constraints, and both continuous and discrete phase shifters.
In particular, we reformulate the original complex problem into a solvable form that can be addressed using the penalty dual decomposition (PDD) method.
Simulation results demonstrate that the proposed design achieves performance close to that of the fully digital (FD) scheme and significantly outperforms other schemes.
Furthermore, leveraging UAV mobility and multi-static cooperation provides crucial spatial degrees of freedom, effectively avoiding WSR degradation under limited transmit power or strict sensing requirements.
\end{abstract}

\vspace{-0.1cm}
\section{Introduction}\vspace{-0.1cm}

The sixth-generation (6G) wireless networks are envisioned to support applications such as autonomous driving and drone navigation, which demand both ubiquitous connectivity and high-precision sensing \cite{9737357,li2024near}. 
Integrated sensing and communication (ISAC) has emerged as a key technology to meet these requirements by jointly exploiting spectrum and hardware resources. 
While conventional cellular-based ISAC suffers from limited coverage and severe self-interference, cell-free architectures provide a promising alternative. 
By distributing sensing tasks across multiple nodes, cell-free multi-static ISAC naturally mitigates the self-interference inherent in mono-static systems, while providing multi-angle observation gains to enhance both communication rates and sensing performance~\cite{10494224}.

Although terrestrial ISAC systems have been extensively researched, the stationary infrastructure restricts their performance, particularly when serving mobile users and tracking dynamic sensing targets. 
To overcome these limitations, unmanned aerial vehicles (UAVs) have been integrated into ISAC systems as flexible aerial platforms. 
By optimizing their trajectories, UAVs can actively maneuver to maintain high-quality links and effective observation angles. 
However, while the existing literature has integrated UAVs into cell-free ISAC with joint trajectory and beamforming design \cite{11028051}, the potential for cooperative multi-static sensing remains under-explored.
Furthermore, most prior works assume fully digital (FD) beamforming, which is often impractical for size, weight, and power (SWaP)-constrained UAVs due to the high cost and energy consumption of dedicated radio frequency (RF) chains. 
Although hybrid digital-analog beamforming \cite{11026097} and other efficient antenna architectures \cite{li2026pinching} have been widely adopted in terrestrial ISAC systems to balance performance and hardware costs, its application in multi-UAV cell-free ISAC networks has received limited attention.

Motivated by these gaps, this paper investigates a joint hybrid digital-analog beamforming and trajectory design for multi-UAV-enabled cell-free multi-static ISAC systems. Specifically, we formulate a weighted sum-rate (WSR) maximization problem for the joint design of UAV trajectories and hybrid beamformers, while considering multi-static sensing, kinematic requirements, and both continuous and discrete phase shifters.
To handle the non-convexity and coupling variables, an efficient algorithm is developed based on the penalty dual decomposition (PDD) framework. 
Simulation results demonstrate that the proposed design closely approaches the FD scheme and significantly outperforms other benchmarks. 
In particular, while fixed-trajectory or bi-static schemes often suffer from severe performance degradation or even infeasibility under limited power budgets or strict sensing constraints, our approach maintains high WSR by leveraging UAV mobility and multi-static cooperation.\vspace{-0.1cm}


\section{System Model}\vspace{-0.1cm}

\begin{figure}[!ht]
    \centering\vspace{-0.3cm}
    \includegraphics[width=0.5\textwidth]{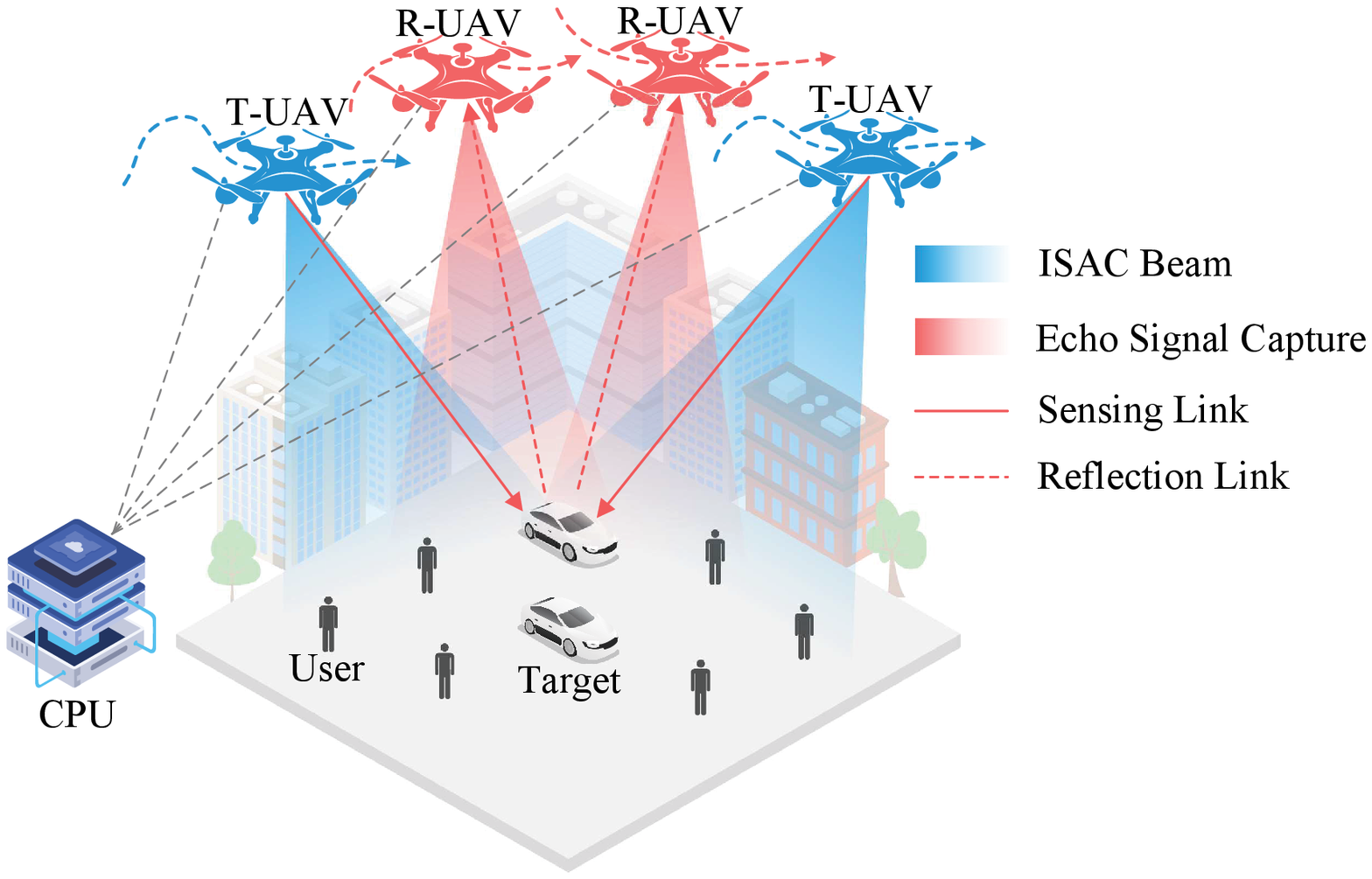}
    \caption{Illustration of the multi-UAV-enabled cell-free multi-static ISAC System.}\vspace{-0.3cm}
    \label{fig_1}
\end{figure}

As shown in Fig.~\ref{fig_1}, consider a multi-UAV-enabled cell-free multi-static ISAC system consisting of $M_t = |\mathcal{M}_t|$  Tx UAVs  and $M_r = |\mathcal{M}_r|$ Rx UAVs. 
A central processing unit (CPU) is deployed to coordinate the operations of all UAVs.
Each Tx UAV is equipped with an $N_t$-antenna uniform linear array (ULA) and a fully-connected hybrid beamforming architecture with $N_{\mathrm{RF}}$ RF chains, where $1 \le N_{\mathrm{RF}} < N_t$. 
Similarly, each Rx UAV is equipped with an $N_r$-antenna ULA for echo signal reception.
In this system, the Tx UAVs cooperatively serve $K = |\mathcal{K}|$ single-antenna ground users and detect a point target.
The set of all signal streams is denoted by $\mathcal{I} = \mathcal{K} \cup \mathcal{S}$, where $\mathcal{S}$ denotes the set of sensing streams with $S = |\mathcal{S}|$ ($K+S \leq M_t N_{\mathrm{RF}}$), and $\mathcal{M} = \mathcal{M}_t \cup \mathcal{M}_r$ denotes the set of all UAVs.
All UAVs are assumed to fly at a constant altitude $H$ over a total duration $T$ in a rectangular area of $L_x \times L_y$, where the users and the target are distributed on the ground.
The UAVs move through this area, and the horizontal position of UAV $m \in \mathcal{M}$ at time slot $t$ is denoted by $\mathbf{q}_m(t) = [q^x_m (t),  q^y_m(t)]^T \in \mathbb{R}^2$. 
Let $\mathbf{g}_k(t) = [g^x_k(t), g^y_k(t)]^T \in \mathbb{R}^2$ and $\mathbf{t}_s(t) = [t^x_s(t), t^y_s(t)]^T \in \mathbb{R}^2$ denote the coordinates of user $k$ and target $s$ at time slot $t$, respectively, which are assumed to be constant and known during the mission period.

\subsection{Signal Model}

Let $\mathbf{F}_{m_t}(t) \in \mathbb{C}^{N_t \times N_{\mathrm{RF}}}$ denote the analog beamformer for UAV $m_t$ at time slot $t$. 
We define $\mathbf{w}_{{m_t},k}(t) \in \mathbb{C}^{N_{\mathrm{RF}} \times 1}$ and $\mathbf{w}_{{m_t},s}(t) \in \mathbb{C}^{N_{\mathrm{RF}} \times 1}$ as the digital beamformers for communication and the sensing waveform, respectively. 
Hence, the transmit signal of UAV ${m_t}$ is expressed as:
\begin{align}
\mathbf{x}_{m_t}(t)
&\!=\! 
\!\sum_{k\in\mathcal{K}}\! \mathbf{F}_{m_t}(t)\mathbf{w}_{{m_t},k}(t)x_k(t)
\!+\!\!
\!\sum_{s\in\mathcal{S}}\! \mathbf{F}_{m_t}(t)\mathbf{w}_{{m_t},s}(t)x_s(t) \nonumber \\
&\!=\!
\!\sum_{i\in\mathcal{I}}\! \mathbf{F}_{m_t}(t)\mathbf{w}_{{m_t},i}(t)x_i(t)
.
\label{eq:composite_signal}
\end{align}
Let $x_k(t) \sim \mathcal{CN}(0, 1)$ be the random communication symbol for user $k \in \mathcal{K}$, and $x_s(t)$ be the dedicated sensing symbol for target $s \in \mathcal{S}$ with unit power. 
These symbols are assumed to satisfy $\mathbb{E}[x_i(t) x_j^*(t)] = 0$ for $\forall i \neq j$.

For the hybrid beamforming architecture, each element of the analog beamformer $\mathbf{F}_{m_t}(t)$ must satisfy $[\mathbf{F}_{m_t}(t)]_{i,j} \in \mathcal{F}$, where the feasible set $\mathcal{F}$ represents either continuous phase shifters, i.e., $\mathcal{F}_C \triangleq \{ e^{\jmath \theta} \mid \theta \in [0,2\pi] \}$, or discrete phase shifters quantized with $\kappa$-bit resolution, i.e., $\mathcal{F}_D \triangleq \{ e^{\jmath \frac{2\pi m}{2^\kappa}} \mid m \in \{0, \dots, 2^\kappa - 1\} \}$.

\subsection{Cooperative Communication Model}

Due to the high altitude of the UAVs, assume the air-to-ground links are characterized by line-of-sight (LoS) propagation. 
At time slot $t$, the channel between Tx UAV ${m_t}$ and user $k$ is modeled as $\mathbf{h}_{{m_t},k}(t) = \frac{\sqrt{\beta_0}}{d_{{m_t},k}(t)} \mathbf{a}_{{m_t},k}(t)$, where $d_{{m_t},k}(t) = \sqrt{\|\mathbf{q}_{{m_t}}(t) - \mathbf{g}_k(t)\|_2^2 + H^2}$ and $\beta_0$ is the reference path loss. 
The steering vector is $\mathbf{a}_{{m_t},k}(t) = [1, \dots, e^{\jmath \frac{2\pi d}{\lambda}(N_t-1)\cos \theta_{{m_t},k}(t)}]^T$ with AoD $\theta_{{m_t},k}(t) = \arccos (H/d_{{m_t},k}(t))$, where $\lambda = c/f_c$ is the carrier wavelength.

Let $\bar{\mathbf{q}}(t) = [\mathbf{q}_1^T(t), \dots, \mathbf{q}_{|\mathcal{M}|}^T(t)]^T \in \mathbb{R}^{2M \times 1}$, $\bar{\mathbf{h}}_k\big(\bar{\mathbf{q}}(t)\big) = [\mathbf{h}_{1,k}^T\big(\bar{\mathbf{q}}(t)\big), \ldots, \mathbf{h}_{M_t,k}^T\big(\bar{\mathbf{q}}(t)\big)]^T \in \mathbb{C}^{M_t N_t \times 1}$, $\bar{\mathbf{F}}(t) = \mathrm{blkdiag}(\mathbf{F}_1(t), \dots, \mathbf{F}_{M_t}(t)) \in \mathbb{C}^{M_t N_t \times M_t N_{\mathrm{RF}} } $, and $\bar{\mathbf{w}}_i(t) = [\mathbf{w}_{1,i}^T(t), \dots, \mathbf{w}_{M_t,i}^T(t)]^T \in \mathbb{C}^{M_t N_{\mathrm{RF}} \times 1 }  $. 
The received signal at UE $k$ at time slot $t$ can be expressed as:
\begin{align}
&y^{(c)}_k(t) = \sum_{{m_t} \in \mathcal{M}_t} \mathbf{h}_{{m_t},k}^H\big(\bar{\mathbf{q}}(t)\big) \mathbf{x}_{m_t}(t) + n_k(t) \nonumber \\
&= \bar{\mathbf{h}}_k^H\big(\bar{\mathbf{q}}(t)\big) \bar{\mathbf{F}}(t) \bar{\mathbf{w}}_k(t) x_k(t) 
\!+ \!\!\!\! \sum_{k' \neq k } \! \bar{\mathbf{h}}_k^H\big(\bar{\mathbf{q}}(t)\big) \bar{\mathbf{F}}(t)\bar{\mathbf{w}}_{k'}(t) x_{k'}(t) \nonumber \\
& + \sum_{s \in \mathcal{S}} \bar{\mathbf{h}}_k^H\big(\bar{\mathbf{q}}(t)\big) \bar{\mathbf{F}}(t)  \bar{\mathbf{w}}_s(t) x_s(t)
+ n_k(t), 
\label{eq:rx_signal}
\end{align}
where $n_k(t) \sim \mathcal{CN}(0, \sigma_k^2)$ denotes the additive receiver noise at UE $k$. 
Based on the signal model in \eqref{eq:rx_signal}, the signal-to-interference-plus-noise ratio (SINR) for user $k$ at time slot $t$ is given by:
\begin{align}
\gamma^{(c)}_k (t) =\! \frac{|\bar{\mathbf{h}}_k^H(t) \bar{\mathbf{F}}(t) \bar{\mathbf{w}}_k(t)|^2}
{\sum\limits_{k' \neq k} \!\!\! |\bar{\mathbf{h}}_k^H(t) \bar{\mathbf{F}}(t) \bar{\mathbf{w}}_{k'}(t)|^2 
\!\!+\!\!\! \sum\limits_{s \in \mathcal{S}}\!\! |\bar{\mathbf{h}}_k^H(t) \bar{\mathbf{F}}(t) \bar{\mathbf{w}}_s(t)|^2\!\! + \! \sigma_k^2}.
\label{eq:comm_sinr}
\end{align}
where the trajectory vector $\bar{\mathbf{q}}(t)$ in the right-hand side is omitted for brevity.
The achievable rate for user $k$ is $\mathrm{R}_k^{(c)}(t) = \log_2(1 + \gamma^{(c)}_k(t))$.

\subsection{Sensing Model}

The channel matrix between Tx UAV $m_t$ and Rx UAV $m_r$ associated with target $s$ at time slot $t$ is modeled as:
\begin{equation}
\label{eq:G_mtmr_q}
\mathbf{G}_{m_r,m_t}\big(\bar{\mathbf{q}}(t)\big)
\!=\! \alpha_s \sqrt{\beta_{m_r,m_t}(t)}\mathbf{a}_{m_r,s}(t)\mathbf{a}_{m_t,s}(t)^H,
\end{equation}
where $\mathbf{a}_{m_t,s}(t) \in \mathbb{C}^{N_t \times 1}$, $\mathbf{a}_{m_r,s}(t) \in \mathbb{C}^{N_r \times 1}$ and $\alpha_s \sim \mathcal{CN}(0,1)$ are the transmit and receive steering vectors and the normalized bi-static radar cross-section (RCS), respectively. 
According to the radar range equation \cite{10494224}, the channel gain is defined as $\beta_{m_r,m_t}(t) = \frac{\lambda^2 \sigma_{\text{RCS}}^2}{(4\pi)^3} d_{m_t,s}^{-2}(t) d_{m_r,s}^{-2}(t)$, where $\sigma_{\text{RCS}}^2$ denotes the variance of the target's bi-static RCS.

Following the Swerling-I model, when the target's velocity is relatively low, $\alpha_s$ remains constant throughout the coherent processing interval (CPI) and any phase-shift associated with the reflection path is incorporated into the unknown RCS $\alpha_s$.
By leveraging the downlink symbols and the channels between UAVs known a priori from the CPU, the resulting echo signal received at Rx UAV $m_r$ is characterized as:\vspace{-0.1cm}
\begin{equation}
\label{eq:yr_mt}
\mathbf{y}^{(s)}_{m_r}(t)
= \sum\nolimits_{m_t\in\mathcal{M}_t} \mathbf{G}_{m_r,m_t}\big(\bar{\mathbf{q}}(t)\big) \mathbf{x}_{m_t}(t) + \mathbf{n}_{m_r}(t),\vspace{-0.1cm}
\end{equation}
where $\mathbf{n}_{m_r} \sim \mathcal{CN}(\mathbf{0}, \sigma_{n}^2 \mathbf{I}_{N_r})$ denotes the noise. 
Accordingly, the total received echo signal $\bar{\mathbf{y}}^{(s)}(t) \in \mathbb{C}^{M_r N_r \times 1}$ at time slot $t$ is formulated as:
\begin{equation}
\label{eq:compact_model}
\bar{\mathbf{y}}^{(s)}(t) = \alpha_s \bar{\mathbf{A}}\big(\bar{\mathbf{q}}(t)\big) \bar{\mathbf{x}}(t) + \bar{\mathbf{n}}(t),
\end{equation}
where $\bar{\mathbf{x}}(t) = [ \mathbf{x}_{1}^T(t), \dots, \mathbf{x}_{M_t}^T(t) ]^T \in \mathbb{C}^{M_t N_t \times 1}$,
$\bar{\mathbf{n}}(t) = [ \mathbf{n}_{1}^T(t), \dots, \mathbf{n}_{M_r}^T(t) ]^T \in \mathbb{C}^{M_r N_r \times 1}$ with $\bar{\mathbf{n}}(t) \sim \mathcal{CN}(\mathbf{0}, \sigma_n^2 \mathbf{I}_{M_r N_r})$,
and 
$
\bar{\mathbf{A}}\big(\bar{\mathbf{q}}(t)\big)
=\sqrt{\frac{\lambda^2 \sigma_{\text{RCS}}^2}{(4\pi)^3 }}\bar{\mathbf{a}}_{r}(t) \bar{\mathbf{a}}_{t}(t)^H \in \mathbb{C}^{M_r N_r \times M_t N_t} 
$
with 
$
\bar{\mathbf{a}}_r(t) = \big[ \frac{\mathbf{a}_{1,s}^T(t)}{d_{1,s}(t)}, \dots, \frac{\mathbf{a}_{M_r,s}^T(t)}{d_{M_r,s}(t)} \big]^T
$, 
$
\bar{\mathbf{a}}_t(t) = \big[ \frac{\mathbf{a}_{1,s}^T(t)}{d_{1,s}(t)},  \dots,\allowbreak \frac{\mathbf{a}_{M_t,s}^T(t)}{d_{M_t,s}(t)} \big]^T
$.

In the considered cell-free ISAC architecture, all UAVs are synchronized in time and frequency under CPU coordination. 
Substituting the optimal matched filter $\mathbf{u}^\star(t) = \frac{1}{\sigma_n^2} \bar{\mathbf{a}}_r(t)$ \cite{11190074} into the sensing signal-to-noise ratio (SNR) and taking the expectation $\mathbb{E}[|\alpha_s|^2]=1$, the SNR is simplified to:
\begin{align}
\label{eq:Rayleigh}
&\gamma^{(s)}(\bar{\mathbf{q}}(t),\bar{\mathbf{F}}(t),\bar{\mathbf{W}}(t)) \nonumber \\
&= \frac{(\mathbf{u}^\star\!(t))^H \bar{\mathbf{A}}\big(\bar{\mathbf{q}}(t)\big) \bar{\mathbf{F}}(t) \bar{\mathbf{W}}(t) \bar{\mathbf{W}}^H\!(t) \bar{\mathbf{F}}^H\!(t) \bar{\mathbf{A}}^H\!\big(\bar{\mathbf{q}}(t)\big) \mathbf{u}^\star\!(t)}{\big(\mathbf{u}^\star\!(t)\big)^H (\sigma_n^2 \mathbf{I}_{M_r N_r})\mathbf{u}^\star\!(t)}  \nonumber \\
&=\frac{\frac{\lambda^2 \sigma_{\text{RCS}}^2}{(4\pi)^3 } \|\bar{\mathbf{a}}_r(t)\|_2^2 \|\bar{\mathbf{a}}_t^H(t) \bar{\mathbf{F}}(t) \bar{\mathbf{W}}(t)\|_2^2}{\sigma_n^2} \\
&= \frac{  \big\|  \bar{\mathbf{A}}\big(\bar{\mathbf{q}}(t)\big)\bar{\mathbf{F}}(t) \bar{\mathbf{W}}(t) \big\|_F^2}{ \sigma_{n}^2},
\end{align} 
where $\bar{\mathbf{W}}(t) = [ \bar{\mathbf{w}}_1(t), \ldots, \bar{\mathbf{w}}_{|\mathcal{I}|}(t)]$.

\subsection{Problem Formulation}
 
In this section, we aim to maximize the downlink WSR by optimizing the UAV trajectories and the analog-digital beamformers. 
For ease of notation, let $\mathcal{Q} = \{\bar{\mathbf{q}}(t)\}_{t=1}^{T}$, $\mathcal{F} = \{\bar{\mathbf{F}}(t)\}_{t=1}^{T}$, and $\mathcal{W} = \{\bar{\mathbf{W}}(t)\}_{t=1}^{T}$.
The resulting WSR hybrid beamforming optimization problem can be given as:\vspace{-0.2cm}
\begin{subequations}\label{P1}
\begin{align}
(\mathbb{P}_1)&\max_{\mathcal{Q},\mathcal{F},\mathcal{W}} \quad  \sum\nolimits_{t=1}^T\!\sum\nolimits_{k=1}^K\!\omega_k \mathrm{R}_k^{(c)}\big(\bar{\mathbf{q}}(t),\bar{\mathbf{F}}(t),\bar{\mathbf{W}}(t)\big) \label{P1_obj}\\[-0.05cm]
\text{s.t.}\quad 
& \sum\nolimits_{i\in\mathcal{I}}\|\mathbf{F}_{m_t}(t) \mathbf{w}_{m_t,i}(t)\|_2^2  \le P_{m_t}, \ \forall m_t, t, \label{P1_power} \\[-0.05cm]
& \gamma^{(s)}\big(\bar{\mathbf{q}}(t),\bar{\mathbf{F}}(t),\bar{\mathbf{W}}(t)\big) \geq \gamma_s, \ \forall  t, \label{P1_SNRs} \\[-0.05cm]
& [\mathbf{F}_{m_t}(t)]_{i,j} \in \mathcal{F}, \ \forall m_t, i, j  \label{P1_F}\\[-0.05cm]
& \mathbf{q}_m(1) = \mathbf{q}_m^I,\  \forall m, \label{P1_Position1} \\[-0.05cm]
&\mathbf{q}_m(T) = \mathbf{q}_m^F, \ \forall m, \label{P1_PositionT}\\
& \|\mathbf q_m(t)-\mathbf q_{m'}(t)\|_2^2 \ge d_{min}^2, \ \forall m \neq m', t,\label{P1_dist}\\[-0.05cm]
& \|\mathbf q_m(t+1)-\mathbf q_m(t)\|_2^2 \le (v_{max}\delta_t)^2, \ \forall m, t.  \label{P1_velocity} 
\end{align}\vspace{-0.6cm}
\end{subequations}

\noindent Constraint \eqref{P1_power} limits the transmit power of each UAV to a maximum budget $P_{m_t}$, while \eqref{P1_SNRs} ensures the sensing SNR exceeds the threshold $\gamma_s$ for reliable target detection. 
The constraint inherent to analog phase shifters is enforced by \eqref{P1_F}. 
Constraints \eqref{P1_Position1}-\eqref{P1_velocity} guarantee flight safety and mobility constraints. 
Constraints \eqref{P1_Position1} and \eqref{P1_PositionT} fix the initial and final coordinates, while \eqref{P1_dist} and \eqref{P1_velocity} impose the collision avoidance distance $d_{min}$ and the maximum velocity $v_{max}$, respectively.

\section{Joint Hybrid Beamforming and Trajectory Design}

\subsection{Transformation of Problem \(\mathbb{P}_1\)}

We first introduce the auxiliary variables $\mathcal{P} = \{\mathbf{P}(t)\}_{t=1}^{T}$, $\mathcal{V} = \{\mathbf{V}(t)\}_{t=1}^{T}$, and $\mathcal{Z} = \{\mathbf{Z}(t)\}_{t=1}^{T}$ as follow:%
\begin{align}
\mathbf{P}(t) &= \bar{\mathbf{H}}^H\big(\bar{\mathbf{q}}(t)\big)\bar{\mathbf{F}}(t)\bar{\mathbf{W}}(t), \label{aux:P}\\
\mathbf{V}(t) &= \bar{\mathbf{F}}(t)\bar{\mathbf{W}}(t), \label{aux:V}\\
\mathbf{Z}(t) &= \bar{\mathbf{A}}\big(\bar{\mathbf{q}}(t)\big)\bar{\mathbf{F}}(t)\bar{\mathbf{W}}(t), \label{aux:Z}
\end{align}
where $\bar{\mathbf{H}}\big(\bar{\mathbf{q}}(t)\big) = [\bar{\mathbf{h}}_1(t), \ldots, \bar{\mathbf{h}}_K(t)]$.
For notational brevity, we use $\bar{\mathbf{H}}(t)$ and $\bar{\mathbf{A}}(t)$ to represent $\bar{\mathbf{H}}\big(\bar{\mathbf{q}}(t)\big)$ and $\bar{\mathbf{A}}\big(\bar{\mathbf{q}}(t)\big)$, respectively, in the subsequent derivations where the trajectory $\bar{\mathbf{q}}(t)$ is held fixed.
With the introduction of these auxiliary variables, the constraints \eqref{P1_power} and \eqref{P1_SNRs} can be equivalently reformulated as:
\begin{align}
&\sum_{i \in \mathcal{I}} \|\mathbf{v}_{m_t,i}(t)\|_2^2 \le P_{m_t}, \ \forall m_t , t, \label{cons:power_V}\\
& \frac{\big\| \mathbf{Z}(t) \big\|_F^2}{ \sigma_{n}^2} \ge \gamma_s, \ \forall t. \label{cons:snr_Z}
\end{align}
where $\mathbf{v}_{m_t,i}(t) = \mathbf{F}_{m_t}(t) \mathbf{w}_{m_t,i}(t)$.

Employing Penalty Dual Decomposition (PDD) \cite{9120361}, the Augmented Lagrangian (AL) problem of $\mathbb{P}_1$ is formulated as:
\begin{align}
&(\mathbb{P}_2)  \quad  \max_{\Omega} \quad
\mathcal{L}_A(\Omega)=\sum_{t=1}^{T}\sum_{k=1}^K \omega_k \mathrm{R}_k^{(c)}\big(\mathbf{P}(t)\big) \nonumber\\
&-\sum_{t=1}^{T}\frac{1}{2\rho(t)}\big\|\mathbf{P}(t) \!- \!\bar{\mathbf{H}}^H\big(\bar{\mathbf{q}}(t)\big)\bar{\mathbf{F}}(t)\bar{\mathbf{W}}(t)  + \rho(t)\mathbf{U}(t)\big\|_F^2 \nonumber\\
&-\sum_{t=1}^{T}\frac{1}{2\rho(t)}\big\|\mathbf{V}(t)\! -\! \bar{\mathbf{F}}(t)\bar{\mathbf{W}}(t) + \rho(t)\mathbf{Y}(t)\big\|_F^2 \nonumber\\
&-\sum_{t=1}^{T}\frac{1}{2\rho(t)}\big\|\mathbf{Z}(t)\! -\! \bar{\mathbf{A}}\big(\bar{\mathbf{q}}(t)\big)\bar{\mathbf{F}}(t)\bar{\mathbf{W}}(t) + \rho(t)\mathbf{T}(t)\big\|_F^2 \label{P2_obj} \\
&\text{s.t.}\quad  \eqref{P1_F} ,\ \eqref{P1_Position1} ,\ \eqref{P1_PositionT} ,\ \eqref{P1_dist} ,\ \eqref{P1_velocity} ,\  \eqref{cons:power_V} ,\ \eqref{cons:snr_Z}. \nonumber
\end{align}
where $\Omega = \{ \bar{\mathbf{q}}(t), \bar{\mathbf{F}}(t), \bar{\mathbf{W}}(t), \mathbf{P}(t), \mathbf{V}(t), \mathbf{Z}(t)\}_{\forall t}$ denotes the set of optimization variables, $\{\rho(t)\}_{t=1}^{T}$ is the penalty parameters and $\{\mathbf{U}(t)\}_{t=1}^{T}$, $\{ \mathbf{Y}(t)\}_{t=1}^{T}$, $ \{\mathbf{T}(t)\}_{t=1}^{T}$ denote the Lagrange multipliers associated with the equality constraints \eqref{aux:P}-\eqref{aux:Z}, respectively.

\subsection{Solving the AL Problem $\mathbb{P}_2$}\label{SolveP2}

To solve the problem $\mathbb{P}_2$ in the $\ell$-th outer iteration, we adopt the BSUM \cite{120891009} method. 
This approach updates each block of variables in the $\tau$-th inner iteration with provable convergence.

\subsubsection{Optimization of $\mathcal{W}$}

Given $\mathcal{Q}^{(\ell,\tau)}$, $\mathcal{P}^{(\ell,\tau)}$, $\mathcal{F}^{(\ell,\tau)}$, $\mathcal{V}^{(\ell,\tau)}$ and $\mathcal{Z}^{(\ell,\tau)}$, optimizing $\mathcal{W}^{(\ell,\tau)}$ can be naturally decomposed into $T$ independent unconstrained subproblems:
\begin{align}\label{P_W}
\min_{\bar{\mathbf{W}}(t)} \quad
&\left\| \big(\bar{\mathbf{H}}^{(\ell,\tau)}(t)\big)^H \bar{\mathbf{F}}^{(\ell,\tau)}(t) \bar{\mathbf{W}}(t)- \mathbf{\Gamma}^{(\ell,\tau)}(t)   \right\|_F^2 \nonumber \\[-0.05cm]
+ &\left\| \bar{\mathbf{F}}^{(\ell,\tau)}(t) \bar{\mathbf{W}}(t) - \mathbf{\Upsilon}^{(\ell,\tau)}(t)  \right\|_F^2 \nonumber \\[-0.05cm]
+  &\left\| \bar{\mathbf{A}}^{(\ell,\tau)}(t) \bar{\mathbf{F}}^{(\ell,\tau)}(t) \bar{\mathbf{W}}(t) - \mathbf{\Lambda}^{(\ell,\tau)}(t)   \right\|_F^2.
\end{align}
where
$\mathbf{\Gamma}^{(\ell,\tau)}(t) = \mathbf{P}^{(\ell,\tau)}\!(t) + \rho^{(\ell)}\!(t) \mathbf{U}^{(\ell)}\!(t)$, 
$\mathbf{\Upsilon}^{(\ell,\tau)}(t)= \mathbf{V}^{(\ell,\tau)}\!(t) + \rho^{(\ell)}\!(t) \mathbf{Y}^{(\ell)}\!(t)$, 
$\mathbf{\Lambda}^{(\ell,\tau)}(t) = \mathbf{Z}^{(\ell,\tau)}\!(t) + \rho^{(\ell)}\!(t) \mathbf{T}^{(\ell)}\!(t)$.

Since \eqref{P_W} is a convex problem, its closed-form solution is derived by setting its first-order derivative to zero: 
\begin{equation}
\label{eq:W_update}
\bar{\mathbf{W}}^\star\!(t) \!=\! \big((\bar{\mathbf{F}}^{(\ell,\tau)}\!(t))^H \mathbf{\Phi}^{(\ell,\tau)}\!(t) \bar{\mathbf{F}}^{(\ell,\tau)}\!(t)\big)^\dagger \! \big(\bar{\mathbf{F}}^{(\ell,\tau)}\!(t)\big)^H \! \mathbf{\Xi}^{(\ell,\tau)}\!(t),
\end{equation}
where 
$\mathbf{\Phi}^{(\ell,\tau)}\!(t)= \mathbf{I}_{M_t N_t} +  \big(\bar{\mathbf{A}}^{(\ell,\tau)}\!(t)\big)^H \bar{\mathbf{A}}^{(\ell,\tau)}\!(t) + \bar{\mathbf{H}}^{(\ell,\tau)}\!(t) \allowbreak \cdot \big(\bar{\mathbf{H}}^{(\ell,\tau)}\!(t)\big)^H$, 
$\mathbf{\Xi}^{(\ell,\tau)}\!(t) = \bar{\mathbf{H}}^{(\ell,\tau)}\!(t) \mathbf{\Gamma}^{(\ell,\tau)}(t) + \mathbf{\Upsilon}^{(\ell,\tau)}(t)+ \big(\bar{\mathbf{A}}^{(\ell,\tau)}\!(t)\big)^H \mathbf{\Lambda}^{(\ell,\tau)}(t)$.

\subsubsection{Optimization of $\mathcal{F}$}

With given $\mathcal{Q}^{(\ell,\tau)}$, $\mathcal{P}^{(\ell,\tau)}$, $\mathcal{W}^{(\ell,\tau+1)}$, $\mathcal{V}^{(\ell,\tau)}$ and $\mathcal{Z}^{(\ell,\tau)}$, the optimization of the block-diagonal matrix $\bar{\mathbf{F}}$ can be decomposed into $T$ independent subproblems:%
\begin{subequations}\label{P_F_Trace}
\begin{align} 
\min_{\bar{\mathbf{F}}(t)} \quad & \operatorname{Tr}\big( \bar{\mathbf{F}}^H(t) \mathbf{B}^{(\ell,\tau)}(t) \bar{\mathbf{F}}(t) \mathbf{D}^{(\ell,\tau)}(t) \big) \nonumber \\
&- 2 \Re \big\{ \operatorname{Tr}\big(\bar{\mathbf{F}}^H(t) \mathbf{C}^{(\ell,\tau)}(t)\big) \big\} \\
\text{s.t.} \quad & [\bar{\mathbf{F}}(t)]_{p,q} \in \mathcal{F}, \quad \forall (p,q) \in \Theta, \label{P1_F_modulus} \\
& [\bar{\mathbf{F}}(t)]_{p,q} = 0, \quad \forall (p,q) \notin \Theta, \label{P1_F_sparsity}
\end{align}
\end{subequations}
where $\mathbf{B}^{(\ell,\tau)}(t) =\mathbf{I}_{M_t N_t} + \big(\bar{\mathbf{A}}^{(\ell,\tau)}(t)\big)^H \bar{\mathbf{A}}^{(\ell,\tau)}(t) + \bar{\mathbf{H}}^{(\ell,\tau)}(t) \big(\bar{\mathbf{H}}^{(\ell,\tau)}(t)\big)^H$,
$\mathbf{C}^{(\ell,\tau)}(t) = \big[ \bar{\mathbf{H}}^{(\ell,\tau)}(t) \mathbf{\Gamma}^{(\ell,\tau)}(t) + \big(\bar{\mathbf{A}}^{(\ell,\tau)}(t)\big)^H \mathbf{\Lambda}^{(\ell,\tau)}(t) + \mathbf{\Upsilon}^{(\ell,\tau)}(t) \big] \big(\bar{\mathbf{W}}^{(\ell,\tau+1)}(t)\big)^H$, 
$\mathbf{D}^{(\ell,\tau)}(t) \allowbreak = \bar{\mathbf{W}}^{(\ell,\tau+1)}(t) \big(\bar{\mathbf{W}}^{(\ell,\tau+1)}(t)\big)^H$
and $\Theta$ denotes the block-diagonal indices.
By ignoring constant terms, we employ the Block Coordinate Descent (BCD) method \cite{10839437} to solve  \eqref{P_F_Trace}:
\begin{equation}\label{F_BCD}
\max_{[\bar{\mathbf{F}}(t)]_{p,q}} \Re { [b^{(\ell,\tau,\mu)}(t)]_{p,q}^* [\bar{\mathbf{F}}(t)]_{p,q} }, \quad \text{s.t.} \quad \eqref{P1_F_modulus} ,\ \eqref{P1_F_sparsity},
\end{equation}
where $ [b^{(\ell,\tau,\mu)}(t)]_{p,q} = [\mathbf B^{(\ell,\tau)}(t)]_{p,p} [\bar{\mathbf{F}}^{(\mu)}(t)]_{p,q} [\mathbf D^{(\ell,\tau)}(t)]_{q,q} \allowbreak - [\mathbf B^{(\ell,\tau)}(t) \bar{\mathbf{F}}^{(\mu)}(t) \mathbf D^{(\ell,\tau)}(t)]_{p,q} + [\mathbf C^{(\ell,\tau)}(t)]_{p,q}$.
For continuous phase shifters, the optimal element-wise update is given by:
\begin{equation} \label{eq:CPS_sol}
[\bar{\mathbf{F}}^{(\mu+1)}(t)]_{p,q} = \exp\big(\jmath \angle [b^{(\ell,\tau,\mu)}(t)]_{p,q}\big).
\end{equation}
For discrete phase shifters, the optimal solution is given by:
\begin{equation} \label{eq:DPS_sol}
[\bar{\mathbf{F}}^{(\mu+1)}(t)]_{p,q} = \arg \max_{[\bar{\mathbf{F}}(t)]_{p,q} \in \mathcal{F}_D} \Re \big\{ [b^{(\ell,\tau,\mu)}(t)]_{p,q}^* [\bar{\mathbf{F}}(t)]_{p,q} \big\}.
\end{equation}
Elements $(p,q) \notin \Theta$ remain zero throughout the iterations.
We iteratively update each element until convergence.

\subsubsection{Update of $\mathcal{V}$}

Given  $\mathcal{F}^{(\ell,\tau+1)}$, and $\mathcal{W}^{(\ell,\tau+1)}$, the optimization of $\mathcal{V}$ at time slot $t$ can be naturally decomposed into $M_t$ independent subproblems. 
\begin{align} \label{P_V}
\min_{\mathbf{V}_{m_t}(t) } \quad   \sum_{i \in \mathcal{I}} \big\| \mathbf{v}_{m_t,i}(t) - \mathbf{x}_{m_t,i}^{(\ell,\tau)}(t)\big\|_2^2,  
\quad \text{s.t.}\quad  \eqref{cons:power_V}. 
\end{align}
where $\mathbf{x}_{m_t,i}^{(\ell,\tau)}(t) = \mathbf{F}_{m_t}^{(\ell,\tau+1)}(t) \mathbf{w}_{m_t,i}^{(\ell,\tau+1)}(t) - \rho^{(\ell)}(t) \mathbf{Y}_{m_t,i}^{(\ell)}(t)$.
To satisfy the constraint \eqref{cons:power_V}, the solution for $i\in \mathcal{I}$, $m_t \in \mathcal{M}_t$ is further projected onto a Frobenius-norm ball:
\begin{equation}\label{eq:V_update}
\mathbf{v}_{m_t,i}^{\star}(t) = \mathbf{x}_{m_t,i}^{(\ell,\tau)}(t) \cdot \min \left\{ \frac{\sqrt{P_{m_t}}}{\sqrt{\sum_{j \in \mathcal{I}} \|\mathbf{x}_{m_t,j}^{(\ell,\tau)}(t)\|_2^2}}, 1 \right\}.
\end{equation}

\subsubsection{Update of $\mathcal{Z}$}

Given $\mathcal{Q}^{(\ell,\tau)}$, $\mathcal{F}^{(\ell,\tau+1)}$, and $\mathcal{W}^{(\ell,\tau+1)}$, the optimization of $\mathcal{Z}$ at time slot $t$ can be expressed as:
\begin{equation}\label{P_Z}
\min_{\mathbf{Z}(t)} \quad  \|\mathbf{Z}(t) - \mathbf{\Omega}^{(\ell,\tau)}(t) \|_F^2, 
\quad \text{s.t.} \quad  \eqref{cons:snr_Z},
\end{equation}
where $\mathbf{\Omega}^{(\ell,\tau)}(t) = \bar{\mathbf{A}}^{(\ell,\tau)}(t) \bar{\mathbf{F}}^{(\ell,\tau+1)}(t) \bar{\mathbf{W}}^{(\ell,\tau+1)}(t) - \rho^{(\ell)}(t) \allowbreak \cdot \mathbf{T}^{(\ell)}(t)$.

Since $\| \mathbf{Z}(t) \|_F^2$ is convex, we employ the Majorization-Minimization (MM)  method to linearize the non-convex constraint \eqref{cons:snr_Z} using its first-order Taylor expansion around the given inner iterate  $\mathbf{Z}^{(\mu)}(t)$ as a global lower bound.
\begin{align}\label{prob:Z_surrogate}
\min_{\mathbf{Z}(t)} \ & \|\mathbf{Z}(t) - \mathbf{\Omega}^{(\ell,\tau)}(t) \|_F^2, \ \! \text{s.t.}  \ \Re\{ \!\langle \mathbf{Z}^{(\mu)}(t), \mathbf{Z}(t) \rangle \! \} \! \ge\!  \gamma_s',
\end{align}
where $\gamma_s' = \frac{1}{2}(\gamma_s \sigma_{n}^2  + \|\mathbf{Z}^{(\mu)}(t)\|_F^2)$. 
The optimal solution to \eqref{prob:Z_surrogate} is given by:
\begin{equation}\label{eq:Z_update}
\mathbf{Z}^{(\mu+1)}\!(t) \!=\! \mathbf{\Omega}^{(\ell,\tau)}\!(t) + \frac{[\gamma_s' \!-\! \Re \{ \langle \mathbf{Z}^{(\mu)}\!(t), \mathbf{\Omega}^{(\ell,\tau)}\!(t) \rangle \}]^+}{\|\mathbf{Z}^{(\mu)}\!(t)\|_F^2} \mathbf{Z}^{(\mu)}\!(t),
\end{equation}
where $[x]^+ = \max(0, x)$.
We iteratively update $\mathbf{Z}^{(\mu)}(t)$ until convergence to obtain the final $\mathbf{Z}^{(\ell,\tau+1)}(t)$.

\subsubsection{Optimization of $\mathcal{P}$}

Given $\mathcal{Q}^{(\ell,\tau)}$, $\mathcal{F}^{(\ell,\tau+1)}$, and $\mathcal{W}^{(\ell,\tau+1)}$, the optimization of $\mathcal{P}$ at time $t$ reduces to:
\begin{equation} \label{P_P}
\max_{\mathbf{P}(t)} \ \sum_{k=1}^K \omega_k \mathrm{R}_k^{(c)}\big(\mathbf{p}_k(t)\big)
- \frac{1}{2\rho^{(\ell)}(t)}\big\|\mathbf{P}(t) - \mathbf{\Psi}^{(\ell,\tau)}(t) \big\|_F^2,
\end{equation}
where $\mathbf{\Psi}^{(\ell,\tau)}(t) = \big(\bar{\mathbf{H}}^{(\ell,\tau)}(t)\big)^H \bar{\mathbf{F}}^{(\ell,\tau+1)}(t)\bar{\mathbf{W}}^{(\ell,\tau+1)}(t) - \rho^{(\ell)}(t) \mathbf{U}^{(\ell)}(t)$.

To handle the non-convex $\mathrm{R}_k^{(c)}\big(\mathbf{p}_k(t)\big)$, we construct a concave function at each MM iteration $\mu$, which is proved to be a global lower bound for the original function \cite{10839437}.
$ \mathrm{R}_k^{(c)}(\mathbf{p}_k(t)) \geq c_k^{(\mu)}(t) \sum_{i=1}^I |p_{k,i}(t)|^2 + \sum_{i=1}^I \Re\{ \big(d_{k,i}^{(\mu)}(t)\big)^* p_{k,i}(t)\} +  \text{const}$,
where  $d_{k,i}^{(\mu)} = \mathbf{1}\{i=k\} 2p_{k,k}^{(\mu)} / (\alpha_k^{(\mu)} \ln 2)$ and $c_k^{(\mu)} = -|p_{k,k}^{(\mu)}|^2 / (\alpha_k^{(\mu)} \beta_k^{(\mu)} \ln 2)$
in which $\alpha_k^{(\mu)}(t) = \sum_{i\neq k} |p_{k,i}^{(\mu)}(t)|^2+\sigma_k^2$, 
$\beta_k^{(\mu)}(t) = \alpha_k^{(\mu)}(t) + |p_{k,k}^{(\mu)}(t)|^2$
and $p_{k,i}(t) = \bar{\mathbf{h}}_k^H(t)\bar{\mathbf{F}}(t)\bar{\mathbf{w}}_i(t)$.
Substituting this approximation into \eqref{P_P} results in a strictly concave quadratic problem, whose closed-form solution is given by:
\begin{equation}\label{eq:p_update}
p_{k,i}^{(\mu+1)}(t) = \frac{\omega_k d_{k,i}^{(\mu)}(t) + \big(\rho^{(\ell)}(t)\big)^{-1} [\mathbf{\Psi}^{(\ell,\tau)}(t)]_{k,i} }{ \big(\rho^{(\ell)}(t)\big)^{-1} - 2\omega_k c_{k}^{(\mu)}(t)}.
\end{equation}
We iteratively update each element until convergence.

\subsubsection{Update of $\mathcal{Q}$}

Given $\mathcal{P}^{(\ell,\tau+1)}$, $\mathcal{Z}^{(\ell,\tau+1)}$, $\mathcal{F}^{(\ell,\tau+1)}$, and $\mathcal{W}^{(\ell,\tau+1)}$, the problem for optimizing $\mathcal{Q}$ reduces to:
\begin{align}\label{P_q}
\min_{\mathcal{Q}} \quad \sum_{t=1}^T  L_q\big(\bar{\mathbf{q}}(t)\big) \quad
\text{s.t.} \quad  \eqref{P1_Position1}, \eqref{P1_PositionT}, \eqref{P1_dist}, \eqref{P1_velocity}, 
\end{align}
where  $L_q\big(\bar{\mathbf{q}}(t)\big) = \big\| \big[\bar{\mathbf{H}}^{(\ell,\tau)}\big(\bar{\mathbf{q}}(t)\big) \big]^H \bar{\mathbf{F}}^{(\ell,\tau+1)}(t)\bar{\mathbf{W}}^{(\ell,\tau+1)}(t)- \mathbf{\Gamma}^{(\ell,\tau+1)}(t) \big\|_F^2 
 + \big\| \bar{\mathbf{A}}^{(\ell,\tau)}\big(\bar{\mathbf{q}}(t)\big) \bar{\mathbf{F}}^{(\ell,\tau+1)}(t)\bar{\mathbf{W}}^{(\ell,\tau+1)}(t) - \mathbf{\Lambda}^{(\ell,\tau+1)}(t) \big\|_F^2$.

To tackle the non-convexity of \eqref{P_q}, we employ the Successive Convex Approximation (SCA) technique \cite{120891009}. We linearize $L_q\big(\bar{\mathbf{q}}(t)\big)$ around the given inner iterate $\bar{\mathbf{q}}^{(\mu)}(t)$ using the first-order Taylor expansion:
\begin{equation} \label{eq:f_hat}
    \hat{L}_q\big(\bar{\mathbf{q}}(t)\big) \! =\!  L_q\big(\bar{\mathbf{q}}^{(\mu)}(t)\big) + \nabla_{\bar{\mathbf{q}}(t)} L_q\big(\bar{\mathbf{q}}^{(\mu)}(t)\big)^T \big(\bar{\mathbf{q}}(t) - \bar{\mathbf{q}}^{(\mu)}(t)\big).
\end{equation}
Simultaneously, the non-convex collision avoidance constraint \eqref{P1_dist} is replaced by its first-order lower bound:
\begin{multline}\label{eq:dist_constraint_approx}
2\big(\mathbf{q}_m^{(\mu)}(t) - \mathbf{q}_{m'}^{(\mu)}(t)\big)^T \big(\mathbf{q}_m(t) - \mathbf{q}_{m'}(t)\big) \\
- \|\mathbf{q}_m^{(\mu)}(t) - \mathbf{q}_{m'}^{(\mu)}(t)\|_2^2 \geq d_{\min}^2.
\end{multline}
To ensure the validity of the convex approximation and guarantee convergence, we incorporate a trust-region constraint:
\begin{equation}
 \| \mathbf{q}_m(t) - \mathbf{q}_m^{(\mu)}(t) \|_2^2 \leq \Delta^2, \quad \forall m, t, \label{cons:trust_region}
\end{equation}
Substituting \eqref{eq:f_hat} and \eqref{eq:dist_constraint_approx} into \eqref{P_q}, the optimization at inner SCA iteration $\mu$ is reformulated as the following convex quadratic program:
\begin{equation} \label{eq:SCA_q}
    \min_{\mathcal{Q}} \quad \sum_{t=1}^T \hat{L}_q\big(\bar{\mathbf{q}}(t)\big), \quad \text{s.t.} \quad \eqref{P1_Position1}, \eqref{P1_PositionT}, \eqref{P1_velocity},  \eqref{eq:dist_constraint_approx}, \eqref{cons:trust_region}, 
\end{equation}
which can be efficiently solved using standard optimization tools such as CVX.

\subsection{Overall Solution of Problem $\mathbb{P}_1$}

The PDD-based solver for $\mathbb{P}_1$ is summarized in Algorithm~\ref{alg:P1_Main}. 
To evaluate the convergence, we define the constraint violation as $\mathcal{E}^{(\ell)}(t) =  \max \{ \| \mathbf{r}_P^{(\ell)}(t) \|_{\infty}, \| \mathbf{r}_V^{(\ell)}(t) \|_{\infty}, \| \mathbf{r}_{Z}^{(\ell)}(t) \|_{\infty} \}$, where the primal residuals are given by 
$\mathbf{r}^{(\ell)}_P(t) = \mathbf{P}^{(\ell)}(t) - \big(\bar{\mathbf{H}}^{(\ell)}(t)\big)^H\bar{\mathbf{F}}^{(\ell)}(t)\bar{\mathbf{W}}^{(\ell)}(t), 
\mathbf{r}^{(\ell)}_V(t) = \mathbf{V}^{(\ell)}(t) - \bar{\mathbf{F}}^{(\ell)}(t)\bar{\mathbf{W}}^{(\ell)}(t),
\mathbf{r}^{(\ell)}_{Z}(t) = \mathbf{Z}^{(\ell)}(t) - \bar{\mathbf{A}}^{(\ell)}(t)\bar{\mathbf{F}}^{(\ell)}(t)\bar{\mathbf{W}}^{(\ell)}(t)$. 
In the inner loop, primal variables are updated as in Section~\ref{SolveP2}, while the outer loop updates dual variables or reduces $\rho^{(\ell)}$ to ensure feasibility.
The algorithm terminates when $\max_t \{\mathcal{E}^{(\ell)}(t)\}$ falls below a predefined tolerance $\epsilon_{2}$.
The total complexity of Algorithm~\ref{alg:P1_Main} is:
$
\mathcal{O}\big( I_{out} I_{in} ( T I_1 M_t^4 N_t^2 N_{\mathrm{RF}}^2 + T M_t^3 N_t^2 N_{\mathrm{RF}} + T  I_2 K |\mathcal{I}| M_t^2 N_t N_{\mathrm{RF}}  +T I_3 |\mathcal{I}| M_t N_t M_r N_r   + I_4 (TM_t)^{3.5} ) \big),
$
where $I_{out}$ and $I_{in}$ denote the outer and inner iteration counts, while $I_1, I_2, I_3$, and $I_4$ are the iteration numbers for updating $\mathcal{F}$, $\mathcal{P}$, $\mathcal{Z}$, and $\mathcal{Q}$, respectively.
Under mild conditions, Algorithm~\ref{alg:P1_Main} is guaranteed to converge to the KKT solutions of problem $\mathbb{P}_1$ for the continuous phase shifters~\cite{9120361}.

\begin{algorithm}[!hb]
\caption{PDD-based Joint Optimization for $\mathbb{P}_1$}\label{alg:P1_Main}
\begin{algorithmic}[1]
\STATE \textbf{Initialize}: $\{\mathcal{Q}, \mathcal{F}, \mathcal{W}, \mathcal{P}, \mathcal{V}, \mathcal{Z} \}^{(0)}$, $\{\rho^{(0)}(t)\}^T_{t=1}$, $\xi$, $\epsilon_1$, $\epsilon_2$, $N_{1}$ , $N_{2}$, $\ell$ and $\tau$.
\REPEAT
    \STATE $\ell = \ell + 1$, $\tau = 0$;
    \REPEAT
        \STATE Update $\{\mathcal{Q}, \mathcal{F}, \mathcal{W}, \mathcal{P}, \mathcal{V}, \mathcal{Z} \}^{(\ell,\tau+1)}$.
        \STATE $\tau = \tau + 1$.
    \UNTIL{$|\mathcal{L}_A^{(\ell,\tau)}-\mathcal{L}_A^{(\ell,\tau-1)}|  / | \mathcal{L}_A^{(\ell,\tau)}| \leq \epsilon_{1}$ or $\tau \geq N_{1}^{\max}$.}
   
    \FOR{each time slot $t = 1, \dots, T$}
        \IF{$\mathcal{E}^{(\ell)}(t) \leq \eta$}
            \STATE $\mathbf{U}^{(\ell+1)}(t) = \mathbf{U}^{(\ell)}(t) + (\rho^{(\ell)}(t))^{-1}\mathbf{r}^{(\ell)}_P(t)$.
            \STATE $\mathbf{Y}^{(\ell+1)}(t) = \mathbf{Y}^{(\ell)}(t) + (\rho^{(\ell)}(t))^{-1} \mathbf{r}^{(\ell)}_V(t)$.
            \STATE $\mathbf{T}^{(\ell+1)}(t) = \mathbf{T}^{(\ell)}(t) + (\rho^{(\ell)}(t))^{-1}\mathbf{r}^{(\ell)}_{Z}(t)$.
            \STATE $\rho^{(\ell+1)}(t) = \rho^{(\ell)}(t)$.
        \ELSE
            \STATE $\{ \mathbf{Y}(t),\! \mathbf{U}(t) ,\! \mathbf{T}(t) \}^{(\ell+1)}\!=\!\{ \mathbf{Y}(t),\! \mathbf{U}(t),\! \mathbf{T}(t) \}^{(\ell)}$
            \STATE $\rho^{(\ell+1)}(t) = \xi \rho^{(\ell)}(t)$.
        \ENDIF
    \ENDFOR
\UNTIL{$\max_t \{\mathcal{E}^{(\ell)}(t)\} \le \epsilon_2$ or $\ell \geq N_{2}^{\max}$.}
\end{algorithmic}
\end{algorithm}

\section{Simulation Results}

We evaluate the proposed design in a multi-UAV-enabled cell-free multi-static ISAC system within a $200 \times 200 \text{ m}^2$ area. 
The simulations are conducted in MATLAB 2023b on an Intel Core i7-13700 platform with 32 GB RAM. 
The results are averaged over 50 independent simulations. 
The system consists of $M_t=M_r=2$ UAVs, $K=7$ ground users, and one target, where each UAV employs an $N_t=N_r=16$ ULA while the users and target follow constant-velocity linear trajectories. 
Key parameters include $H=50$ m, $V_{\max}=20$ m/s, $T=30$ s, $\delta_t=1$ s, $\Delta=2$ m, $f_c = 1.9$ GHz, $d=\lambda/2$, $\beta_0=-50$ dB, $\sigma_{\text{RCS}}^2=1$, $\sigma_k^2 = -110$ dBm, and $\sigma_{n}^2=-94$ dBm. 
PDD hyperparameters are set as $\{\rho^{(0)}(t)\}_{t=1}^T = 0.3$, $\xi=0.8$, $\eta=10^{-4}$, $\epsilon_1=10^{-6}$, $\epsilon_2=10^{-8}$, $\omega_k=1/K$, and $N_{1}^{\max}=N_{2}^{\max}=100$.
The initial and final positions of the Tx UAVs are $\mathbf{q}_1^{I/F} = [180, 150]^T/[20,150]^T$ and $\mathbf{q}_2^{I/F} = [20, 50]^T/[180, 50]^T$, while those of the Rx UAVs are $\mathbf{q}_3^{I/F} = [180, 110]^T/[20,110]^T$ and $\mathbf{q}_4^{I/F} = [20, 90]^T/[180, 90]^T$.

To evaluate the performance of our proposed joint design, we consider five schemes, all optimized within the proposed PDD framework to ensure a fair comparison: 
1) ``FD Comm-only'', which optimizes the WSR for FD beamforming without sensing constraints; 
2) ``FD Joint'',which optimizes the WSR for FD beamforming subject to sensing constraints;
3) ``Fixed Traj'', which optimizes only hybrid beamforming with fixed UAV positions;
4) ``MAP'', which is obtained by first solving the FD Joint problem and then performing matrix approximation (MAP) on the digital beamformers;
5) ``Bi-Static'', which involves only the Tx UAV 1 and the Rx UAV 1.

\begin{figure}[!t]
\centering
\includegraphics[width=2.5in]{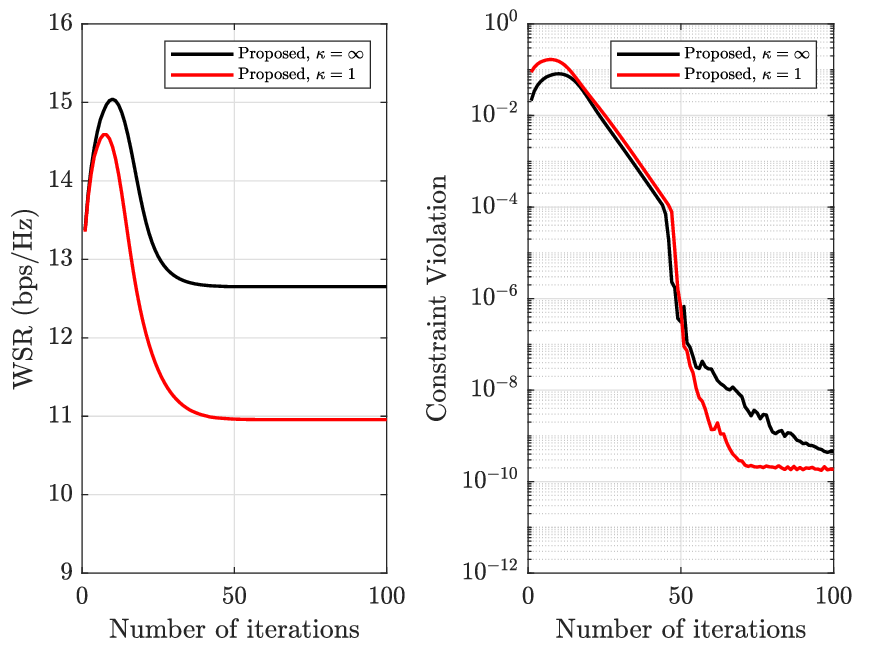}
\caption{Convergence performance of the proposed algorithm. The left and right y-axes represent the WSR and the total constraint violation $\mathcal{E}^{(\ell)}$, respectively ($N_{\mathrm{RF}}=8$, $P_{m_t} = 20$ dBm, $\gamma_s=5$ dB).}
\label{fig_2}
\end{figure}

Fig.~\ref{fig_2} plots the WSR and constraint violation $\mathcal{E}^{(\ell)}$ to illustrate the convergence of the proposed PDD algorithm for both continuous and 1-bit phase shifters.
For both continuous and discrete phase shifters, the WSR initially increases and then decreases to convergence. 
Initially, the WSR increases because the penalty parameter $\rho$ is large, allowing the algorithm to explore the objective space by temporarily violating constraints. 
As $\rho$ decreases, the increasing penalty forces $\mathcal{E}^{(\ell)}$ below $10^{-8}$, ensuring the solution is strictly feasible.
Furthermore, the continuous phase shifter outperforms the discrete case due to the absence of quantization loss.

\begin{figure}[!t]
\centering
\includegraphics[width=2.5in]{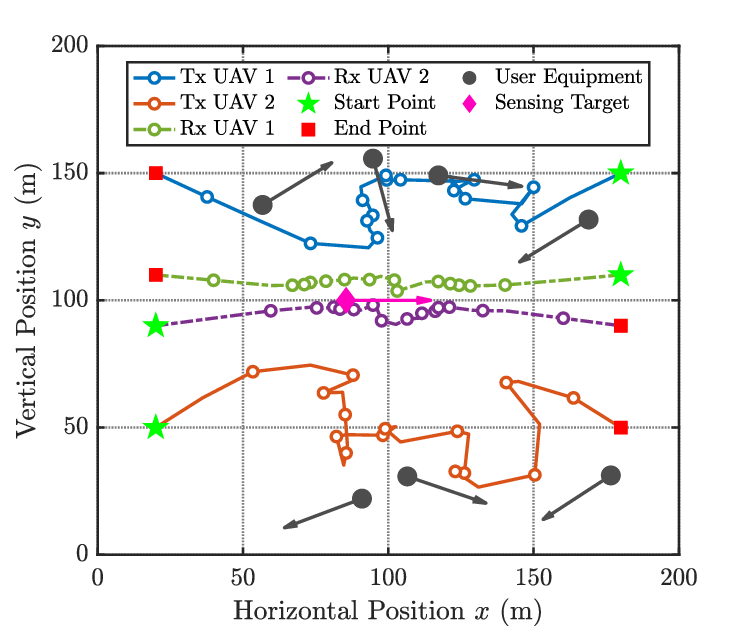}
\caption{Optimized UAV trajectories in cell-Free multi-static ISAC ($N_{\mathrm{RF}}=8$, $\kappa=\infty$, $P_{m_t} = 10$ dBm, $\gamma_s= 20$ dB).}
\label{fig_3}
\end{figure}

Fig.~\ref{fig_3} depicts the optimized trajectories, where markers indicate movement every $2$ time slots. 
Tx UAV $1$ approaches the users near the sensing target to meet both sensing and communication requirements. In contrast, Tx UAV $2$ maintains an intermediate position to strike a trade-off between the communication quality for distant users and the sensing performance.
To satisfy the sensing requirements, the Rx UAVs move closer to the target to capture stronger echoes.

\begin{figure}[!t]
    \centering
    \subfloat[]{
        \includegraphics[width=0.47\columnwidth]{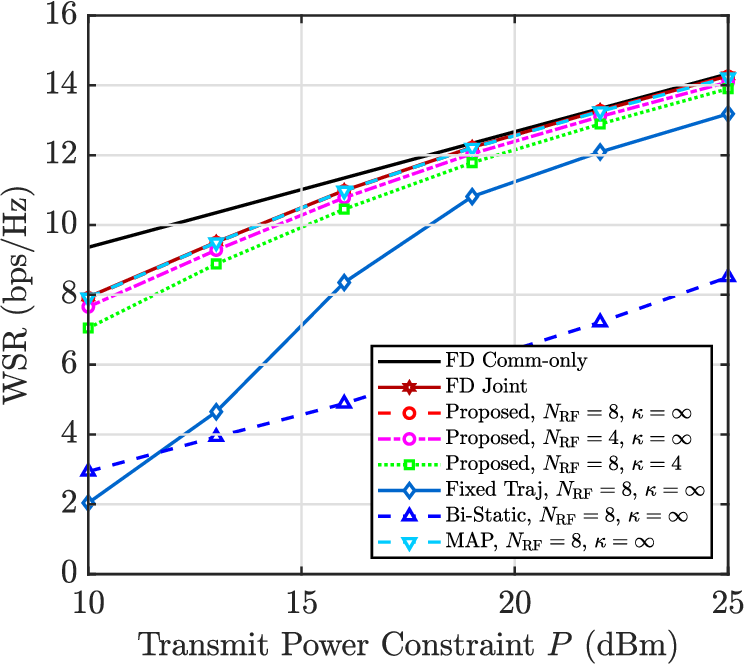}
        \label{fig_4}
    }
    \hfil
    \subfloat[]{
        \includegraphics[width=0.47\columnwidth]{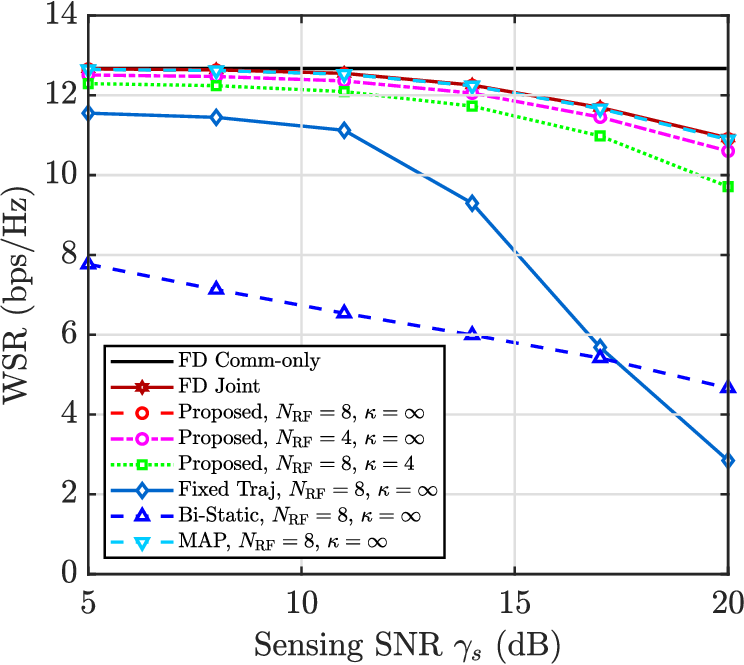}
        \label{fig_5}
    }
    \caption{Performance evaluation of the proposed design versus schemes: (a) WSR versus transmit power budgets $P_{m_t}$ ($\gamma_s = 10$ dB); (b) WSR versus sensing SNR constraints $\gamma_s$ ($P_{m_t} = 20$ dBm).
    }
    \label{fig_performance}
\end{figure}

As shown in Fig.~\ref{fig_4} and Fig.~\ref{fig_5}, our proposed hybrid designs ($N_{\mathrm{RF}}=\{4,8\}$) achieves similar performance as the ``FD Joint'' benchmark, while outperforming other schemes.
Notably, under limited transmit power $P_{m_t}$ or stringent sensing constraints $\gamma_s$, the ``Fixed Traj'' and ``Bi-Static'' baselines suffer from severe WSR degradation due to the lack of spatial flexibility. 
In contrast, our joint design maintains high WSR even under stringent constraints, demonstrating the joint benefit of UAV mobility and multi-static cooperation in managing the communication-sensing trade-off.
Furthermore, discrete phase shifters with $\kappa=4$ bits lead to minimal WSR loss compared to continuous ones, verifying their hardware practicality.

\section{Conclusion}

In this paper, we explored the joint design of hybrid beamforming and UAV trajectories for multi-UAV-enabled cell-free multi-static ISAC systems. 
Accounting for the minimum multi-static sensing requirements, maximum transmit power budgets, phase-shifter constraints, and UAV mobility, we formulated a joint hybrid beamforming and trajectory optimization problem to maximize the WSR.
To handle the highly non-convex problem, we propose a PDD-based iterative algorithm employing MM, and SCA techniques.
Simulation results confirm that the proposed hybrid architecture effectively approaches the performance of the FD scheme, even with discrete phase shifters. 
Furthermore, comparative analyses demonstrate that integrating UAV trajectory design and multi-static cooperation is essential to maintaining a high WSR, especially when transmit power is limited or sensing requirements are strict.

\bibliographystyle{IEEEtran}
\bibliography{IEEEabrv,reference}

\end{document}